\documentstyle[preprint,aps,psfig] {revtex}

\begin{document}

\title{\bf A Lattice Boltzmann method for simulations of liquid-vapor
thermal flows}

\author{Raoyang Zhang and Hudong Chen}
\address{Exa Corporation, 450 Bedford Street, Lexington, MA 02420}

\date{\today}
\maketitle

\begin{abstract}
We present a novel lattice Boltzmann method that has a 
capability of simulating thermodynamic multiphase flows. 
This approach is fully thermodynamically consistent at the macroscopic
level. Using this new method, a liquid-vapor boiling process, including
liquid-vapor formation and coalescence together with
a full coupling of temperature, is simulated for the first time.
\end{abstract}

\newpage


\section{Introduction}

After years of research, the Lattice Boltzmann Methods (LBM) has become
an established numerical approach in computational fluid dynamics (CFD).
Many models and extensions have been formulated that cover a wide range
of complex fluids and flows\cite{syc-doolen}. Furthermore,
LBM has been extended to include
turbulence models that have already had a direct and substantial impact
on engineering applications\cite{exa-papers,molvg,lax,hou}.

Among many desirable LBM features such as simplicity,
parallelizability, and robustness in dealing with 
complex boundary conditions, one recognized advantage is its 
capability of simulating
fluid flows with multiple phases\cite{shan1-2,he1-2,syc-doolen}. 
The core mechanism in LBM modeling of multiphase flows is 
its microscopic level realization of non-ideal
gas equations of state. As a result, at sufficiently low temperature
and proper pressure, liquid-vapor like first order phase 
transitions are spontaneously
generated. There is no need to explicitly tracking
the interfaces between immiscible phases.
Furthermore, unlike static statistical physical 
models\cite{rechl},
LBM also contains the momentum conservation, 
so that bubbles and liquid droplets are formed along with
fluid hydrodynamic processes. The success and simplicity of LBM for
multiphase flows has led to various applications that include simulations
of oil-water mixtures through porous media\cite{roth}, Rayleigh-Taylor
problems\cite{zhang,he1-2}, and many more\cite{syc-doolen}. On the other
hand, there is an crucial missing piece. That is,
so far all the existing multiphase LBM models are limited 
to regimes in which the temperature dynamics 
is either negligible or its effect on
flow is unimportant. This limitation, along with the
overall unavailability in CFD, has prevented
us from dealing with an important class of flows, namely multiphase flows
involving strong couplings with thermodynamics. Specific examples of such
type of flows range from the common water boiling processes to thermal
nuclear reactor applications. Thus
from both fundamental and practical point of views, extensions
of the existing CFD and LBM methods to simulation of 
thermal multiphase flows is extremely important.

The LBM is originally evolved from lattice gas
models obeying fundamental conservation laws and symmetries\cite{fhp1,zati,qian3,succi}.
Now it has also been shown to be systematically derivable from
the continuum Boltzmann equation\cite{shan-he}. 
The most commonly known lattice Boltzmann equation (LBE) has
the following form
(adopting the lattice units convention in which $\Delta t = \Delta
x = 1$),
\begin{equation}
f_i({\bf x} + {\hat c}_i, t + 1) - f_i({\bf x}, t) = {\cal C}_i,
\label{bolz}
\end{equation}
where time $t$ takes on only positive integer values, and the particle
velocity takes on a finite set of discrete vector values (speeds), 
$\{ {\hat c}_i ; \; \; i = 0, \ldots , b\}$. 
These speeds form links among nodes on a given lattice\cite{fhp1,qian}.
The collision term on the right hand side of eqn.(\ref{bolz}) 
now often uses
the so called Bhatnagar-Gross-Krook (BGK) approximation\cite{bgk,qian},
\begin{equation}
{\cal C}_i = - \frac {f_i - f^{eq}_i} {\tau},
\label{bgk}
\end{equation}
having a single relaxation time parameter, $\tau$. Here, $f^{eq}_i$
is the local equilibrium distribution function that has an appropriately prescribed functional dependence on the local hydrodynamic properties. 
The basic hydrodynamic quantities, such as fluid density 
$\rho$ and velocity ${\bf u}$,
are obtained through simple moment summations,
\begin{eqnarray}
\rho ({\bf x}, t) &=& \sum_i f_i ({\bf x}, t), \nonumber \\
\rho {\bf u} ({\bf x}, t) &=& \sum_i {\hat c}_i f_i ({\bf x},t).
\label{momt}
\end{eqnarray}
In addition, one can also define a fluid temperature $T$ from,
\begin{equation}
\rho \frac {D} {2} T ({\bf x}, t) = \sum_i \frac {1} {2} ({\hat c}_i - {\bf u} 
({\bf x}, t))^2 f_i ({\bf x}, t),
\label{tmt}
\end{equation}
where $D$ is the dimension of the momentum space of the discrete lattice
velocities\cite{fhp1}. It has been theoretically
shown that the hydrodynamic behavior
produced from LBE obeys the Navier-Stokes fluid dynamics
at a long wave-length and low frequency limit\cite{qian}. The resulting
equation of state is that of an ideal gas fluid, namely the pressure
$p$ obeys a linear relation with density and temperature,
\begin{equation}
p = \rho T .
\end{equation}
The kinematic viscosity of the fluid is related to the relaxation parameter
by\cite{fhp1,molvg,qian}
\begin{equation}
\nu = (\tau - \frac {1} {2} ) T .
\end{equation}

LBM has been extended to simulations of multiphase
flows\cite{syc-doolen,shan1-2,he1-2}.
The key step is to introduce an additional term, $\Delta f_i({\bf x},t)$ 
on the right hand side of eqn.(\ref{bolz}), to represent a body-force.
This force term is self-consistently generated by the neighboring distribution
functions around each lattice site, and it does not either violate 
the local mass
conservation nor the global momentum conservation. However,
the local momentum is altered by an amount,
\begin{equation}
{\bf F} ({\bf x}, t) = \sum_i {\hat c}_i \Delta f_i({\bf x},t).
\label{force}
\end{equation}
The appearance of the body force term can be physically, in a
mean-field sense, attributed to a non-local 
interacting potential $U$ among the particles\cite{huang}. 
The existence of such an interacting potential is the essential
mechanism in the non-ideal gas type of fluids.
Hence with a suitable choice of $U$, spontaneous phase
separations can be produced, and one can use it conveniently to numerically
study multiphase flow phenomena. 
Through the years, there have been many progresses in LBM models
for multiphase flows\cite{syc-doolen}. On the other hand, as pointed
out at the beginning, all the existing attempts are limited 
to isothermal (or ``a-thermal'')
situations in which the dynamics of temperature in the fluid is suppressed.
That is, $T$ is either assumed a constant or, at best, a prescribed function
of space (or time).

\section{Multiphase flows with incorporation of thermodynamics}

In this paper, we present an extension of multiphase LBM to include
the full thermodynamic process.

The most natural extension in LBM for thermodynamics has been to introduce
a conserved energy degree of freedom\cite{frank}. This is relatively
straightforward for the ideal gas type of models in which only pointwise collisions are involved
and only kinetic energy is considered. When a sufficient number
of particle speeds is used,
one can theoretically show that LBM leads to the
correct full set of thermohydrodynamic equations
of an ideal-gas fluid\cite{molvg,frank,chris}.
Unfortunately, besides being considerably more expensive
computationally than the isothermal LB models, 
such an approach cannot be easily generalized to
multiphase thermodynamic flows. The most obvious obstacle is
the difficulty in tracking the energy evolution while
maintaining a total energy conservation:
For a non-ideal gas system, the total energy also contains an
interacting energy part that is a
function of the relative positions among the particles. Without a
total energy conservation, a temperature variable cannot be defined
fully self-consistently at the microscopic level. In addition, it has
been shown that, unlike the isothermal models, an LBE with an energy
degree of freedom does not guarantee
a global $H$-theorem\cite{chen1-2}. As a consequence, the
system can exhibit significantly less stability.
Other undesirable features in this direct approach include
1) difficulty in changing Prandtl number value from
unity, unless a substantial generalization to the BGK
collision term is made;
and 2) a rather limited temperature range (with the 
maximal allowable value only about twice the minimal value), 
unless significantly 
more speeds are added\cite{molvg,chris}.
Because of these reasons, the progress in LBM
for thermal multiphase flows has been rather slow.

Here we present a new LBM approach that can essentially avoid
all of the above mentioned drawbacks. The fundamental idea 
can be briefly summarized: First of all, the fluid dynamics part
(i.e., the density and momentum evolution) is represented by a modified
{\it isothermal} LBE, while the energy evolution part is determined by
an additional scalar energy transport equation\cite{Bartoloni}. The latter 
can be solved either via a finite difference scheme or an auxiliary
LBE. Secondly, the coupling of the two parts is through 
a properly defined body force
term in the LBE (and the compression and dissipation terms 
in the energy equation).
As we shall realize below, although conceptually rather
simple, the new model produces the correct full
thermohydrodynamic equations together with a non-ideal 
gas equation of state.

We choose a common isothermal
LBE (e.g., D3Q19,\cite{qian}) as a starting basis.
As discussed earlier, an isothermal LBE model
for fluid density and velocity evolution is considerably simpler 
compared to its energy conserving counterpart.
This is certainly desirable for
doing efficient fluid flow simulations.
Furthermore, the equilibrium distribution in
an isothermal LB model is only a function of fluid density and velocity.
The lack of temperature dynamics in
the equilibrium distribution is the
key for achieving a higher stability in LBE\cite{chen1-2}.
Having these in mind, it is much desirable to introduce 
a macroscopic mechanism to recover thermodynamics. 
Specifically, instead of letting temperature to influence
the equilibrium distributions in an LB system, the thermodynamic effect
is obtained via a temperature-dependent body-force\cite{shan1-2}. 
Because of the ``external'' nature of the coupling, 
the LB system and its equilibrium property remain to be
microscopically isothermal.  Nonetheless, as explained below, 
this alternative way of coupling achieves 
the desired thermodynamics at the macroscopic level.

Ignoring the higher order contributions,
the body-force term can be simply expressed as\cite{martys},
\begin{equation}
\Delta f_i({\bf x},t) = \frac {w_i} {T_0} {\hat c}_i \cdot {\bf F},
({\bf x}, t)
\label{linear}
\end{equation}
where the constant weights $w_i$ and $T_0$ are directly determined
by the LBE model (e.g., D3Q19;, in which $T_0 = 1/3$). 
One can easily verify that this gives rise to
eqn.(\ref{force}). The global momentum conservation is
preserved as long as ${\bf F} ({\bf x}, t)$ is expressed as 
a spatial gradient of a scalar function\cite{qian2},
\begin{equation}
{\bf F} ({\bf x}, t) = - \nabla U ({\bf x}, t).
\label{potl}
\end{equation}
It is straightforward to implement this condition in a discrete
space by proper finite-difference procedures. 
Based on the consideration for
higher order isometry in surface tension, we choose (for D3Q19) the
following specific form,
\begin{equation}
\nabla U ({\bf x}, t) \approx \sum_i
\frac {D} {bc^2_i} {\hat c}_i U({\bf x} + {\hat c}_i, t)
\label{fd}
\end{equation}
With the additional body-force term, one can easily recognize that
the overall effective pressure in the resulting
fluid momentum equation has become,
\begin{equation}
p = \rho T_0 + U
\label{eos1}
\end{equation}
where the first term is a result of the
isothermal LBM. From (\ref{eos1}), 
one can obtain any form of equation of state, 
$p = p(\rho, T)$, simply by making a corresponding choice for $U$,
\begin{equation}
U({\bf x}, t) = p(\rho ({\bf x}, t), T({\bf x}, t)) 
- \rho ({\bf x},t) T_0.
\label{total}
\end{equation}
The above quantity is determined once the local values of 
$\rho ({\bf x}, t)$ and $T({\bf x}, t)$ are provided.
Obviously the resulting fluid is no longer isothermal
if the temperature $T({\bf x}, t)$ varies.
In other words, because of the macroscopic way of coupling,
the resulting fluid dynamics is no longer isothermal. In addition,
with the high flexibility of choosing the equation 
of state, this approach can be applied to simulation
of non-ideal gas fluids and multiphase flows. 
Indeed, we confirmed this basic feature through 
a set of spinodal decomposition tests
based on a Van de Waals gas model (Carnahan-Starling equation of state).
Similar to the other multiphase LBM, 
a spontaneous phase separation process is well observed 
at sufficiently low temperature values.

The evolution of the temperature
$T({\bf x}, t)$ in this new approach
is obtained from solving
a supplemental scalar energy transport equation, 
\begin{equation}
\rho (\partial_t + {\bf u}\cdot \nabla ) e = - p\nabla \cdot {\bf u}
+ \nabla \cdot \kappa \nabla T + \Psi
\label{temp}
\end{equation}
where $e = c_v T$ is the internal energy, and $c_v$ is the specific
heat at constant volume of the fluid. The overall pressure
$p$ is defined by the equation of state (\ref{eos1}), and $\kappa$ is 
the heat conductivity that can be specified flexibly. 
The term $\Psi$ represents the viscous dissipation
of flow and the contribution of surface tension.
The energy evolution equation (\ref{temp}) is a standard macroscopic
description for thermal fluids\cite{huang}.
The computation of an isothermal LB model along
with a scalar energy equation
is considerably less expensive than any microscopic attempts: for it
neither requires many particle speeds nor complicated tracking of the
energy evolution. Moreover, the difficulties in
stability and Prandtl number associated with the original thermal LBM
are not issues in this approach.
Solving a scalar transport equation is rather straightforward. 
There are many finite-difference schemes for
accurately and efficiently solving the
scalar transport equation. In our particular simulations, we have used
an extended Lax-Wendroff scheme\cite{lax}. The combination of eqns.
(\ref{bolz})-(\ref{momt}), (\ref{linear})-(\ref{fd}), and
(\ref{total})-(\ref{temp}) forms the new LBM approach for modeling
multiphase thermodynamic fluid flows. 
The thermal boundary condition can be
realized via standard numerical procedures so that,
\begin{equation}
\kappa {\hat n} \cdot \nabla T |_w = q,
\label{thbc}
\end{equation}
with a prescribed heat flux $q$, that can either be fixed or a 
function of local properties in order to achieve a fixed wall 
temperature. The unit vector ${\hat n}$ denotes the surface 
normal direction.

There is one more important feature in this model worth pointing out.
That is, the new approach avoids the fundamental limitation on 
the temperature range that has constrained 
the other thermal LB models.
Notice the temperature only appears in 
the body-force term in a gradient function form. 
Hence, unlike that for equilibrium distribution functions, 
there is no absolute upper or lower
bound on the temperature values except that it should not change too
rapidly across a given resolution scale. Moreover, there is obviously
no absolute bound on temperature in the energy equation.

Based on the above discussions, one can realize that the 
new LB model generates a fully macroscopically
consistent description for thermodynamic flows involving 
generalized equations of state. Therefore, this approach
offers a convenient and efficient numerical tool for studying
thermal multiphase flow problems.

\section{Simulation of the liquid-vapor boiling process}

In this section we present computational results of a typical
multiphase thermodynamic flow simulation
with the new LB approach. In particular, a liquid-vapor
boiling process involving Rayleigh-Benard like convections, phase
changes, together with a complex temperature dynamics 
is simulated successfully, albeit qualitative.
Although representative of a wide range of important
applications, boiling flow problems have received little success from
CFD in general. Consequently, it is very important that
the new approach can demonstrate such a fundamental capability.

The Rayleigh-Benard convection process has been widely used as a benchmark
for many fluid computations. It is the simplest representation of a
boiling phenomenon in which a complex buoyancy-driven convection process
occurs at various values of the Rayleigh number\cite{chanders}. On
the other hand, most of the boiling processes occur in nature also
involve evolutions of multiple thermodynamic phases. 
That is, besides thermal convection,
the fluid under goes a phase transition process in 
which liquid droplets and vapor bubbles are generated.
The most obvious practical examples include the common
water boiling in a pot. 

We choose a standard Rayleigh-Benard setup, in which both the upper
and the lower solid plates obey no-slip boundary conditions, while
the horizontal boundary condition is periodic. To achieve more stable
and second-order accurate numerical results,
we have also applied the new scheme to the modified LB
discretization formulation of He et al\cite{he1-2}.
As discussed above, the Carnahan-Starling equation
of state is used here for convenience. 
The mean density value $\rho = 1.36 \rho_c$.
The temperature on the upper wall is fixed at $T_u = 0.795 T_c$ while on
the lower wall $T_l = 0.954 T_c$. Here $T_c$ ($= 0.55$ in lattice units
for the choice of the model) is the critical temperature
for formation of two phases\cite{he1-2}. The initial temperature is
set to be linearly distributed between the two plates and is
consistent with their temperature boundary conditions.
The simulation volume is $L \times H = 256 \times 128$ grid points.
The gravity value $g = 5 \times 10^{-6}$ (lattice units)
is used. In order to avoid unnecessary complications, a
weak surface tension effect\cite{zhang,he1-2} is also applied in the
LBM flow simulation, so that the surface tension 
contribution to the energy evolution can thus be neglected.
Specifically, we choose the surface tension
coefficient $\sigma$ to be $0.01$ for these simulations.
The kinematic viscosity and the Prandtl number are set at $\nu = 0.02$
and $Pr = 10$, respectively.
For simplicity, the heat capacity $c_v$ in our multiphase flow is
chosen as a constant ($=1$).
All the other fluid parameters are the same as in He et al\cite{he1-2}.
Based on the choice of these parameter values, the resulting Rayleigh
number is $R_a \sim 3.0 \times 10^5$, which is much higher than the first
threshold ($Ra_c = 1708$) for an onset of convection in the conventional
single-phase Rayleigh-Benard system\cite{chanders}.

The simulation starts from a uniform density distribution with one percent
random fluctuations. To enhance bubble formation, small temperature
fluctuations are added to the equation of state
in the first grid point layer near the bottom solid plate. Due to the
higher temperature, a lower density fluid is been produced that subsequently
leads to small vapor bubbles at the bottom plate,
and then these merge with each
other to form larger size ones. The phase formation process
in the simulation is combined with the convection process in which
the hotter and lighter vapor phase rises while
the colder and heavier liquid phase descends due to gravity.
With the above choice of parameter values, such a full
thermohydrodynamic cycle is seen to be able to 
sustain itself indefinitely.
Figure 1 shows the density distributions at some
representative nondimensionalized (by $\sqrt{H/g}$) times.
One can observe the formation of two streams of bubbles
near the bottom plate. One can also observe that
two pairs of counter rotating convection rolls are locked
between the bubble streams. In addition, the convection rolls are seen
to pinch off the small bubbles from
the bottom plate. As bubbles rise, their sizes are seen to increase
slightly. Since small bubbles move faster than large bubbles,
collision and coalescence often occur among them. 
Figure 2 shows streamlines of the
velocity field at time 39.5, from which one can clearly see the two
pairs of counter rotating convection rolls. 
Figure 3 depicts the corresponding
temperature deviation from the linear distribution 
at the same time. Interestingly, the temperature exhibits a 
non-trivial behavior: Its
value is seen to be relatively lower in the vapor phase domains near
their interfaces. As expected physically, this phenomenon is due
to the $p\nabla \cdot {\bf u}$ term in the energy equation (\ref{temp})
associated with the volume expansion from
water to vapor phases. All the above simulated phenomena are
qualitatively correct for a realistic two phase thermodynamic flow.

We also ran another simulation with the exact same setup as the above,
except that the surface tension is made ten times stronger. As seen in
figure 4, only one pair of counter rotating convection 
rolls is produced at its asymptotic state. Different 
from the standard single phase Bernard convection, we 
see that other intrinsic properties such as the surface tension 
can also alter the thermal convection characteristics in a multiphase flow\cite{chanders}.

\section{Discussions}

In this paper, we present a novel approach that combines a multiphase 
lattice Boltzmann method with a scalar temperature equation. 
The coupling is realized macroscopically via
a self-consistent body-force. The basic formulation is applicable
to both 2D and 3D flow situations.
It is directly verifiable theoretically that new LB model obeys the
correct full thermohydrodynamic equations with a non-ideal gas equation
of state. The incorporation of thermohydrodynamics 
in the new approach allows for simulations of complex multiphase 
flows coupled with temperature dynamics.
Other important features of the new scheme include its simplicity,
efficiency and robustness for simulating thermal multiphase 
flow processes\cite{cip}.
The latter has shown to be extremely difficult with
the other LBM based schemes or the more conventional 
numerical methods\cite{saleski}.
Furthermore, like other LBMs, the new approach can handle
complex physical boundary conditions \cite{exa-papers}. 
All these are highly
desirable for a practical computational model.

We have demonstrated the capability of our new approach for doing
boiling flow simulations. This type of flows has so far not been
successfully handled via other methods. Thus the new approach has opened
a promising opportunity for numerically studying
thermal multiphase flow problems. 
On the other hand, further improvements in the direction of the
new approach are necessary in order for it to become 
a more useful and quantitative computational tool.
Without going into detail, the major remaining issues include: 
1) incorporating more realistic equations of state
instead of the van der Waals type of gas model; 2) a more physical
treatment of the heat capacity and latent heat; 
3) further understanding and modeling of surface 
tension and near interface physics; 
4) further enhancement of numerical stability to achieve 
significantly higher density ratio and lower viscosity; 
and 5) generalization of boundary conditions.
Among all the above tasks, we consider (4) to be the most challenging
one for LBM in general.

\section{Acknowledgments}
The authors wish to thank Drs. Xiaoyi He, YueHong Qian, Ilya Staroselsky,
Adrian Tentner, Steven Orszag, Sauro Succi, Rick Shock and Nick Martys for
their useful discussions. 
This work is supported in part by DOE SBIR grant No. 65016B01-I.

\vfill\eject

\newpage

%
%
\begin{figure*}[htp]
\noindent FIG.~1.
Density distributions at t = 39.5, 41.5, 43.5, and 45.5, respectively. Density ratio between liquid and vapor is 3. Red color represents the liquid phase, 
bule color represents the vapor phase.
\label{denvol}
\end{figure*}

%
%
\begin{figure*}[htp]
\noindent FIG.~2.
Streamline and vector plots of the flow velocity field at t = 39.5.
\label{streamline}
\end{figure*}

%
%
\begin{figure*}[htp]
\noindent FIG.~3.
Temperature deviation ($ T_{dev} = {T-T_{linear} \over T_{top}}$) at t = 39.5. The color range is (-0.1, 0.02).
\label{temperature}
\end{figure*}

%
%
\begin{figure*}[htp]
\noindent FIG.~4.
Density distribution and velocity vector plot with surface tension coefficient
$\sigma = 0.1$.
\label{den_st}
\end{figure*}

\newpage

\begin{figure}
\centerline{\hbox{
\psfig{figure=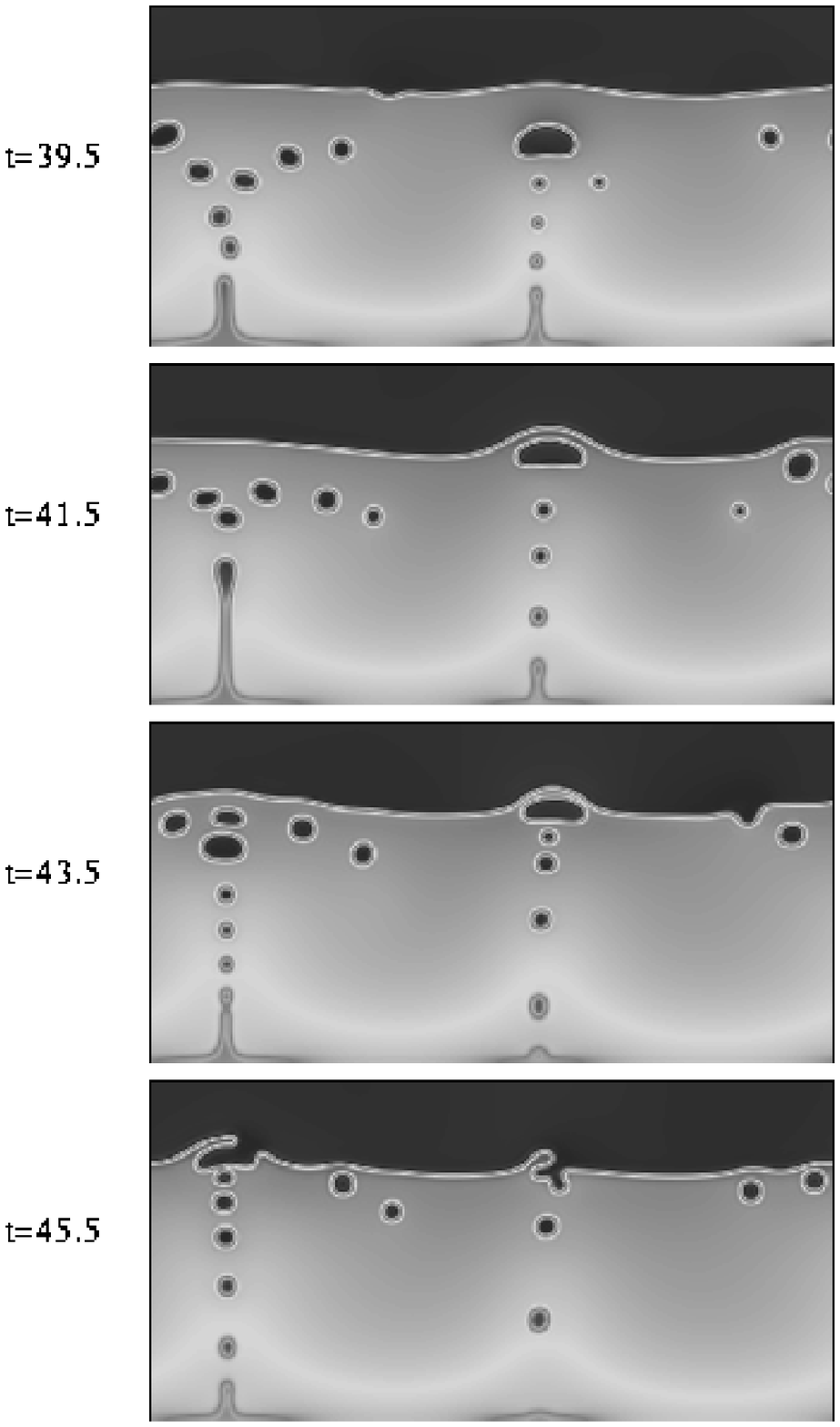,height=9.0in,width=8.0in}
}}
\vskip -2 cm
\caption{}
\end{figure}

\begin{figure}
\centerline{\hbox{
\psfig{figure=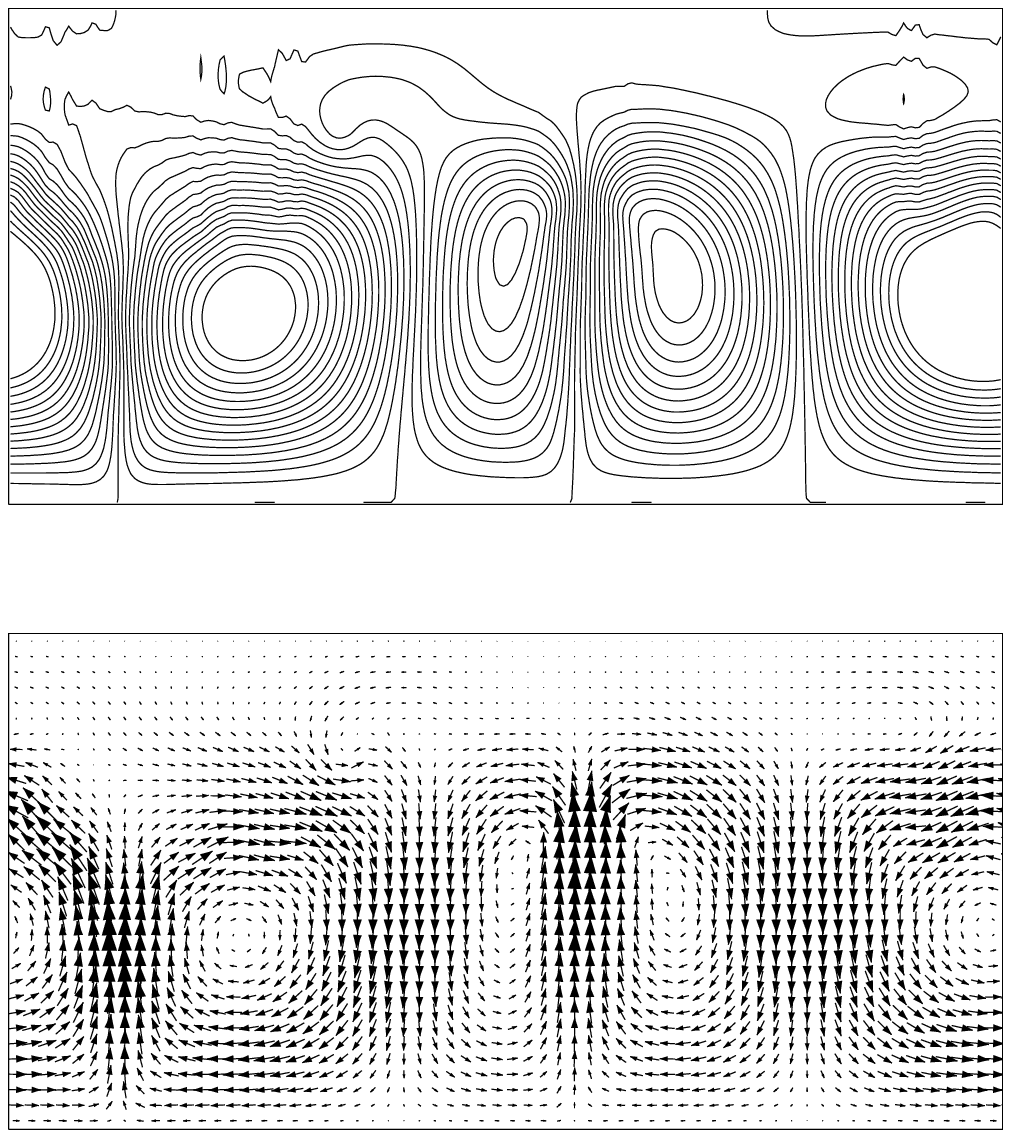,height=4.5in,width=6.5in}
}}
\vskip 6 cm
\caption{}
\end{figure}

\begin{figure}
\centerline{\hbox{
\psfig{figure=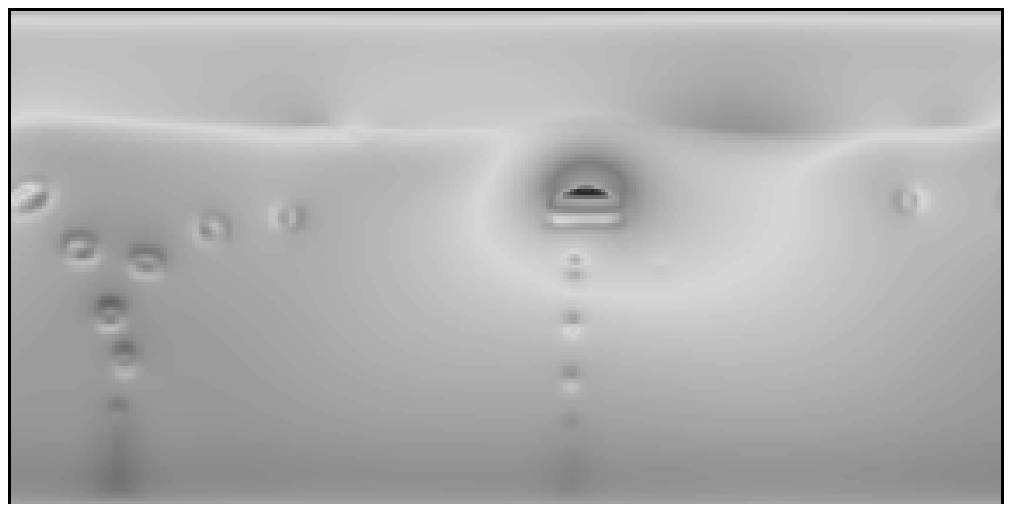,height=9.0in,width=8.0in}
}}
\vskip -2 cm
\caption{}
\end{figure}

\begin{figure}
\centerline{\hbox{
\psfig{figure=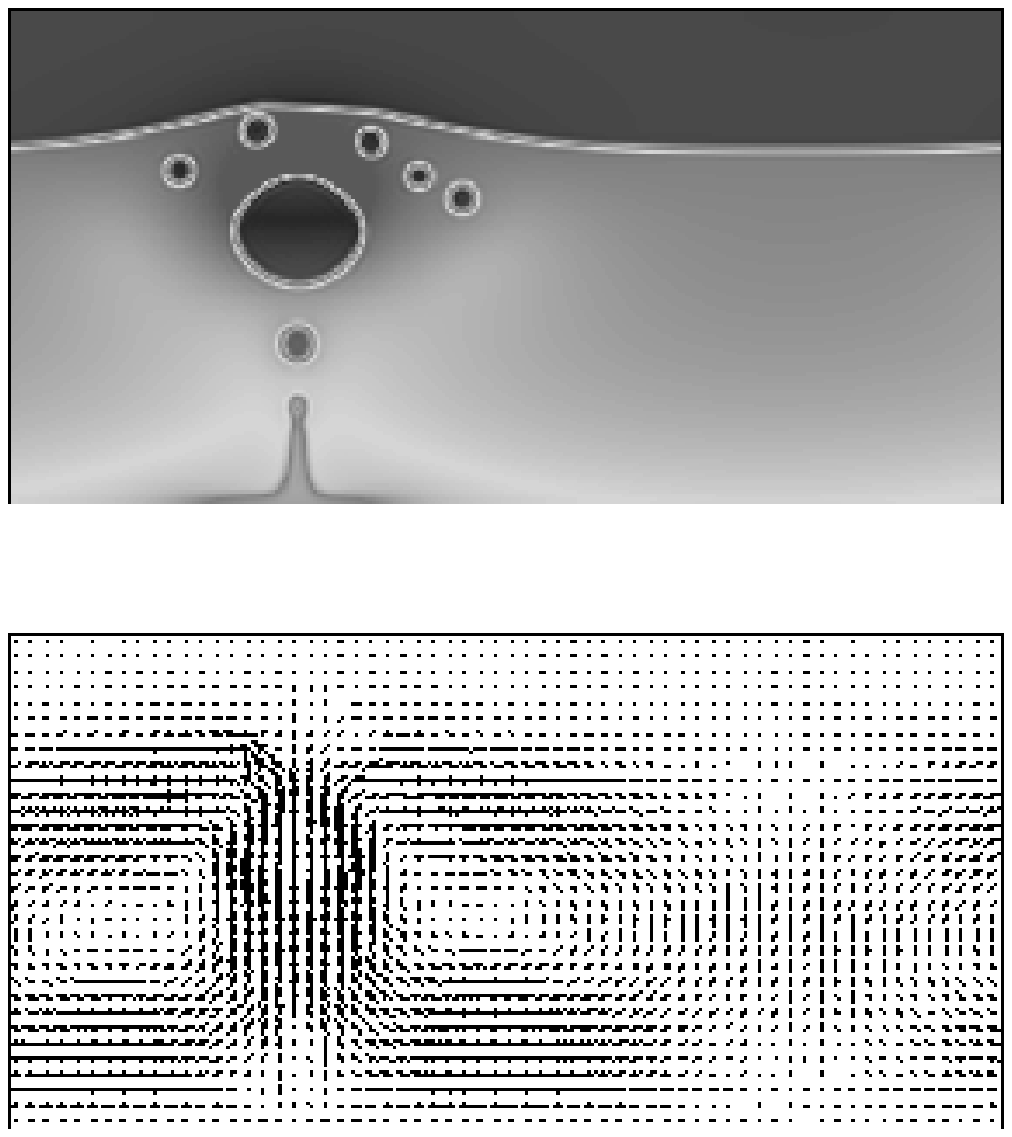,height=9.0in,width=8.0in}
}}
\vskip -2 cm
\caption{}
\end{figure}

\end{document}